
%
%
\input harvmac

\noblackbox

\Title {\vbox{\baselineskip12pt\hbox{BUHEP-94-8}\hbox{hep-ph/9404209}}}
{\vbox{\centerline{A Heavy Top Quark and the
$Zb\bar b$ Vertex}\vskip2pt
\centerline{in Non-Commuting Extended Technicolor}}}


\centerline{R.S. Chivukula, E.H. Simmons and J. Terning\footnote{}
{e-mail addresses: sekhar@abel.bu.edu, simmons@smyrd.bu.edu,
terning@calvin.bu.edu}}
\bigskip
\centerline{Department of Physics}
\centerline{Boston University}
\centerline{590 Commonwealth Ave.}
\centerline{Boston, MA 02215}


\vskip .3in

We explore corrections to electroweak parameters in the context of
Extended Technicolor (ETC) models in which the ETC gauge-boson which
generates the top-quark mass carries weak $SU(2)$ charge.  For $m_t
\sim 150$ GeV there exist potentially large corrections to the $Z$
decay width to $b$-quarks. Interestingly, in contrast to the situation
in ETC models where the gauge-boson which generates the top-quark mass
is a weak $SU(2)$ singlet, the corrections may {\it increase} the
$Z \to b \bar{b}$ branching ratio.

\Date{04/94} 


\def\NPB{{\it Nucl. Phys.} {\bf B}}
\def\PLB{{\it Phys. Lett.}}
\def\PRD{{\it Phys. Rev.} {\bf D}}
\def\PRL{{\it Phys. Rev. Lett.}}

\nref\etczbb{R.~S.~Chivukula,  S.~B.~Selipsky, and E.~H.~Simmons,
\PRL\ {\bf 69} (1992) 575\semi
E~H.~Simmons, R~S.~Chivukula and S~B.~Selipsky, in proceedings of {\it
Beyond the Standard Model III}, S.~Godfrey and P.~Kalyniak, eds.,
World Scientific, Singapore, 1993\semi
N.~Kitazawa \PLB\ {\bf B313} (1993)
395 .}
\nref\strongetc{T.~Appelquist, M.~Einhorn, T.~Takeuchi, and
L~C~R.~Wijewardhana,
{\it Phys. Lett.} {\bf B220} (1989)  223\semi
V.~A.~Miransky and K.~Yamawaki, {\it Mod. Phys. Lett.}
{\bf A4} (1989) 129\semi
K.~Matumoto {\it Prog. Theor. Phys. Lett.} {\bf 81} (1989) 277\semi
V.~A.~Miransky, M.~Tanabashi, and K.~Yamawaki, {\it Phys. Lett.}
{\bf B221} (1989) 177\semi
V.~A.~Miransky, M.~Tanabashi, and K.~Yamawaki, {\it Mod. Phys.
Lett.} {\bf A4} (1989) 1043.}
\nref\lane{R.~S.~Chivukula, K.~Lane, and A.~G.~Cohen, \NPB{\bf 343} (1990)
554.}
\nref\evans{N.~Evans, University of Wales, Swansea preprint SWAT/27,
hep-ph/9403318.}
\nref\walkzbb{R.~S.~Chivukula,  E.~Gates, E.~H.~Simmons, and
J.~Terning, \PLB\ {\bf B311} (1993) 157}
\nref\lepresults{``The Top Mass from LEP \& SLC Electroweak
Measurements'', talk by B. Jacobsen, presented at
workshop on ``QCD and High Energy Hadronic Interactions'', XIXth
Recontres de Moriond, March 19-24, 1994.}
\nref\mtref{``Search for the Top Quark in $p\bar p$ Collisions
at $\sqrt{s}=1.8$ TeV'', S.~Abaschi, {\it et. al.}, The D\O\
Collaboration, submitted to Physical Review Letters.}
\nref\nda{A.~Manohar and H.~Georgi, \NPB{234} (1984) 189.}
\nref\howardbook{ See, for example, H.~Georgi, {\it Weak
Interactions and Modern Particle Theory}, (Benjamin-Cummings, Menlo Park,
1984), p.77. }
\nref\largen{G. 't Hooft, \NPB{\bf 72} {1974} {461}.}
\nref\ununify{H.~Georgi, E.~E.~Jenkins, and E.~H.~Simmons
\PRL\ {\bf 62} (1989) 2789, ERRATUM, ibid. {\bf 63} (1989) 1540
and \NPB {331} (1990) 541.}
\nref\acref{T.~Maskawa and H.~Nakajima, Prog. Theor. Phys. {\bf 52} (1974) 1326
and {\bf 54} (1976) 860\semi
R.~Fukuda and T.~Kugo, \NPB{\bf 117} (1976) 250\semi
K.~Higashijima, \PRD{\bf 29} (1984) 1228\semi
P.~Castorina and S.~Y.~Pi, \PRD{\bf 31} (1985) 411\semi
R.~Casalbuoni, S.~De~Curtis, D.~Dominici, and R.~Gatto, \PLB\ {\bf B150}
(1985) 295\semi
T.~Banks and S.~Raby, \PRD {\bf 14} (1976) 2182\semi
M.~Peskin, in {\it Recent Advances in Field Theory and Statistical
Mechanics}, Les Houches 1982, J.~B.~Zuber and R.~Stora, eds.,
North-Holland, Amsterdam, 1984\semi
A.~Cohen and H.~Georgi, \NPB{\bf 314} (1989) 7.}
\nref\hgsw{H.~Georgi and S.~Weinberg, \PRD{\bf17} (1978) 275.}
\nref\tcew{M.~Golden and L.~Randall, {\it Nucl. Phys.} {\bf B361} (1991) 3\semi
B.~Holdom and J.~Terning, {\it Phys. Lett.} {\bf B247} (1990) 88\semi
M.~Peskin and T.~Takeuchi, {\it Phys. Rev. Lett.} {\bf 65} (1990) 964\semi
A.~Dobado, D.~Espriu, and M.~Herrero, {\it Phys. Lett.} {\bf B253} (1991)
161\semi
M.~Peskin and T.~Takeuchi {\it Phys. Rev.} {\bf D46} (1992) 381\semi
D.~Kennedy, Fermilab preprint {\bf FERMILAB-CONF-93/023-T}.}
\nref\terning{R.~S.~Chivukula, E.~H.~Simmons, and J.~Terning, in preparation.}

\def\st{\sin\theta}
\def\ct{\cos\theta}
\def\sp{\sin\phi}
\def\cp{\cos\phi}

\def\gp{g^\prime}
\def\gq{g_l}
\def\gl{g_h}

\def\su2w{$SU(2)_W$}
\def\gae{\raise-.5ex\vbox{\hbox{$\; >\;$}\vskip-2.9ex\hbox{$\;\sim\;$}}}
\def\lae{\raise-.5ex\vbox{\hbox{$\; <\;$}\vskip-2.9ex\hbox{$\; \sim\;$}}}

\newsec{Introduction}

It has recently been shown that corrections to the $Zb\bar b$ vertex
can place stringent constraints on models of dynamical electroweak
symmetry breaking \etczbb\ . In particular, for extended technicolor
(ETC) models\foot{We confine our attention to models in which the ETC
interactions do not participate directly in electroweak symmetry
breaking, {\it i.e.} we will not consider ``strong-ETC'' models
\strongetc . The fine-tuning required in strong-ETC models will give
rise to additional light scalar degrees of freedom \lane, and the
corrections to the $Zb\bar{b}$ vertex can be much smaller \etczbb
\evans.} in which the ETC gauge-boson giving rise to the top-quark
mass does not carry weak $SU(2)$ charge, it has been found that the
exchange of this gauge-boson alters the $Zb\bar b$ vertex by an amount
proportional to $m_t$.  These corrections decrease the $Z \to b \bar{b}$
branching ratio and are large enough, even in walking technicolor
models \walkzbb , that current LEP data
\lepresults\ appear to exclude such ETC models if
the top mass is 100 GeV or greater (see ref. \mtref).

Finding similar constraints on models in which the ETC gauge-boson
which generates the top-quark mass {\it does} carry weak $SU(2)$
charge is more complicated.  As noted in \etczbb\ both the size and
sign of the effect on the $Zb\bar b$ vertex are model-dependent
because there may be contributions from two different sets of
operators (those involving weak doublet and weak triplet currents).

In this letter, we make a detailed analysis of ETC-related effects on
the $Zb\bar b$ vertex in the context of such ``non-commuting'' models.
We show that the two competing contributions come from the physics of
top-quark mass generation and from weak gauge boson mixing.  We
determine the magnitudes of the effects in terms of the top-quark mass
and the strength of the ETC coupling.  Furthermore, we show that each
of these contributions is of fixed sign -- and that the signs of the
two effects are opposite. Therefore, unlike the case of commuting ETC
models, we cannot make a model-independent statement of the overall
size or sign of the change in the $Z \to b \bar{b}$ branching ratio: the
overall effect may be small and may even {\it increase} the $Z \to b
\bar{b}$ branching ratio.

The starting point of our analysis is the general symmetry-breaking
pattern which non-commuting ETC groups must exhibit in order to
produce phenomenology consistent with both a heavy top-quark and
approximate Cabibbo universality.  A heavy top-quark must receive its
mass from ETC dynamics at low energy scales; if the ETC bosons
responsible for $m_t$ are weak-charged, the weak group $SU(2)_{heavy}$
under which $(t,b)_L$ is a doublet must be embedded in the low-scale
ETC group.  Conversely, the light quarks and leptons cannot be charged
under the low-scale ETC group lest they also receive large
contributions to their masses; hence the weak $SU(2)_{light}$ group
for the light quarks and leptons must be distinct from
$SU(2)_{heavy}$.  To approximately preserve low-energy Cabibbo
universality the two weak $SU(2)$s must break to their diagonal
subgroup before technicolor dynamically breaks the remaining
electroweak symmetry to electromagnetism.  The resulting
symmetry-breaking pattern is\foot{The hypercharge group, $U(1)_Y$, is
embedded partly in the ETC group, and therefore $U(1)' \neq U(1)_Y$.}:
\eqn\lotsa{\eqalign{ETC & \times SU(2)_{light} \times U(1)'\cr
&\downarrow\ \ \ \ \ f \cr
TC \times SU(2)_{heavy} & \times SU(2)_{light} \times U(1)_Y \cr
&\downarrow\ \ \ \ \ u \cr
TC & \times SU(2)_{weak} \times U(1)_Y\cr
&\downarrow\ \ \ \ \ v \cr
TC & \times U(1)_{EM},\cr}}
\smallskip\noindent
where $ETC$ and $TC$ stand for the extended technicolor and
technicolor gauge groups respectively, and $f$, $u$, and $v\approx
246$ GeV are scales of the expectation values of the order parameters
for the three different symmetry breakings ({\it i.e.} the analogs of
$F_\pi$ for chiral symmetry breaking in QCD). Note that, since we are
interested in the physics associated with top-quark mass generation,
only $t_L$, $b_L$ and $t_R$ need transform non-trivially under $ETC$.
Thus $(t,b)_L$ is a doublet under $SU(2)_{heavy}$ but a singlet under
$SU(2)_{light}$, while all other left-handed ordinary fermions have
the opposite $SU(2)$ assignment.

In section 2, we consider how the ETC dynamics responsible for
generating the top-quark mass affects the $Zb\bar b$ vertex.  Section
3 examines the electroweak structure of the model and the impact of
weak gauge-boson mixing on the $Z \to b \bar{b}$ branching ratio.  In
section 4, we give a scenario for a theory with a non-commuting ETC
group.  Section 5 summarizes our conclusions.

\newsec{Effects from ETC dynamics responsible for $m_t$}

One effect on the $Z \to b \bar{b}$ branching ratio comes directly from the
dynamics related to the generation of the mass of the top-quark.  In
ETC models where the gauge-boson whose exchange gives rise to the
top-quark mass carries weak $SU(2)$ quantum numbers, that boson
transforms as a weak doublet.  At energies below the scale $f$ of ETC
breaking, the effective four-fermion operator giving rise to the
top-quark mass is
\eqn\tmasop{- {2\over f^2}\left(\xi\bar\psi_L \gamma^\mu U_L +
{1\over \xi}\bar t_R \gamma^\mu T_R\right) \left(\xi \bar U_L \gamma_\mu
\psi_L + {1\over \xi}\bar T_R \gamma_\mu t_R \right) }
where the left-handed heavy quarks and right-handed technifermions,
$\psi_L = (t,b)_L$ and $T_R = (U,D)_R$, are doublets under
$SU(2)_{heavy}$ while the left-handed technifermions are
$SU(2)_{heavy}$ singlets.  Note that, since $\bar\psi_L \gamma^\mu
U_L$ and $\bar t_R \gamma^\mu T_R$ must transform in the same way, if
$\psi_L$ is a 2 of $SU(2)_{heavy}$ then $T_R$ is a 2* instead.  The
operator \tmasop\ is normalized using the conventional definition of
$f$ with $M_{ETC} = g_{ETC} f/2$ and the conventional factor of
$1/\sqrt{2}$ for an ``off-diagonal'' gauge-boson coupling.  The
parameter $\xi$ in equation \tmasop\ is a model-dependent Clebsch and
is equal to 1 in the class of models outlined in section 4.

The piece of \tmasop\ contributing to $m_t$ is
the product of left-handed and right-handed currents.
Fierzing this into the product of technicolor-singlet
densities gives
\eqn\tmfierz{ {4\over f^2} \left(\bar\psi_L t_R \right)
\left(\bar T_R U_L\right) + h.c.}
When the technifermions condense (we estimate the size of the
condensate using dimensional analysis \nda ,$\langle \bar U U \rangle
\approx 4
\pi v^3$) the top-quark receives a mass
\eqn\tmval{m_t \approx {8 \pi v^3 \over f^2.}}
Hence the relationship between the ETC and TC scales is linear in the
top-quark mass
\eqn\rewrite{{v^2\over f^2} \approx {m_t \over {8\pi v}}.}

The purely left-handed piece of operator \tmasop\
affects the $Zb\bar{b}$ vertex \etczbb .  To see how this occurs,
we Fierz the left-left operator
into the product of technicolor-singlet currents
\eqn\vertex{- {2\xi^2\over f^2} \left( \bar\psi_L \gamma^\mu \psi_L
\right) \left( \bar U_L \gamma_\mu U_L \right) .}
Adopting an effective chiral Lagrangian description appropriate below the
technicolor chiral symmetry breaking scale, we may replace the technifermion
current by a sigma-model current \howardbook :
  \eqn\interpolate{ \left({\bar T}_L \gamma_\mu \tau^a T_L \right) =
	{v^2 \over 2}Tr\left(\Sigma^\dagger\tau^a iD_\mu\Sigma\right)\ ,}
where $\Sigma = \exp{(2i{\tilde\pi}/v)}$ transforms as
$\Sigma \rightarrow L\Sigma R^\dagger$ under $SU(2)_L \times SU(2)_R$.
To evaluate $D^\mu\Sigma$, recall that since $\psi_L$ is a 2 and $T_R$
is a 2* of $SU(2)_L$
\eqn\derivs{ \eqalign{
&D_\mu \psi_L = \del_\mu \psi_L + i {e \over{\st \ct}}
Z_\mu ({1 \over 2}\tau_3 - s^2 Q) \psi_L   + . . .\cr
&D_\mu T_R = \del_\mu T_R + i {e \over{\st \ct}} Z_\mu ({1 \over
2}(-\tau_3^*) -s^2 Q) T_R + . . .\cr}}
so that
\eqn\derivsig{
D_\mu \Sigma = \del_\mu \Sigma - {{i e}\over{\st \ct}} Z_\mu
\left({1\over 2} \Sigma \tau_3^*  + \sin^2\theta[\Sigma,Q]\right) + ...}
Going to unitary gauge ($\Sigma = 1$) we find
\eqn\vertun{\bar U_L \gamma_\mu U_L = - {e \over {\st \ct}} {v^2 \over
4} Z_\mu .}
Combining this with \vertex\ gives the ETC-induced coupling between
the $Z$ and the t-b doublet
\eqn\cuop{\xi^2 {e \over \st \ct} {v^2 \over  2 f^2} \left[\bar\psi_L
\gamma^\mu Z_\mu \psi_L\right]~ .}

This additional coupling between the $Z$ and left-handed $b$ quarks
changes the tree-level $Z b_L \bar b_L$ coupling by an amount\foot{ In
eqn. \tmval\ we have scaled from QCD, implicitly assuming that
technicolor is a non-walking $SU(N)$ gauge group with $N=3$. We may
use large-$N$ results \largen\ to estimate that the size of the effect
is proportional to $\sqrt{N}$.} \eqn\change{\delta g_L = -{e \over{
\st \ct}} {\xi^2 v^2 \over { 2 f^2}} \approx - {\xi^2 \over 4} {e
\over {\st \ct}} {m_t \over {4 \pi v}} .} Since the tree-level $Z b_L
\bar b_L$ coupling is also negative, the ETC-induced change tends to
{\bf increase} the coupling -- and thereby the $Z$ decay width to $b$
quarks ($\Gamma_b$) : \eqn\width{\eqalign{{\delta \Gamma_b \over
{\Gamma_b}} & = 2 { g_L \delta g_L \over {g_L^2 + g_R^2}} \approx {2
\delta g_L \over g_L} \cr & \approx {\xi^2 \over {1 - {2\over 3}
\sin^2 \theta}} {m_t \over {4\pi v}}\cr & \approx + 5.6\%\ \xi^2\
\left({m_t\over {150 {\rm GeV}}}\right) .\cr}}
While this result is of the same magnitude as the change in $\Gamma_b$
induced by top-mass generation in models where the ETC and weak
$SU(2)$ groups commute \etczbb , because the technifermions transform
as a $2^*$ the result is of the opposite sign.

In order to compare these results to experiment, we consider the
ratio
\eqn\defr{
R_b = {\Gamma_b \over \Gamma_{hadrons}}\ .}
Both oblique effects and
the leading QCD corrections cancel in this ratio along with some
experimental systematic effects. We find that eqn. \width\ results in
a change to $R_b$ of
\eqn\shiftr{
{\delta R_b \over R_b} \approx +4.4\% \xi^2 \left({m_t
\over 150{\rm GeV}} \right).}
This is particularly interesting since
recent results \lepresults\ from the LEP Electroweak Working
Group find that the measured value of $R_b$
\eqn\data{R_b = 0.2207 \pm 0.0009 \pm 0.0020\ \ , }
is {\it larger} than the Standard Model prediction of approximately
0.216 for a 150 GeV top-quark mass.  Furthermore, the Standard Model
prediction {\it decreases} with increasing top mass.

\newsec{Effects from weak gauge boson mixing}

Next we explore the effect of the weak gauge boson mixing in
non-commuting ETC models on the $Z \to b \bar{b}$ branching ratio. The
electroweak group structure $SU(2)_{heavy}\times SU(2)_{light} \times
U(1)_Y$ implies the existence of an extra set of massive $W$ and $Z$
gauge bosons.  Hence the light $W$ and $Z$ mass eigenstates will
differ from the standard $W$ and Z.

The electroweak symmetry breaking pattern is similar to that of the
un-unified standard model \ununify , and our notation is chosen to
facilitate this comparison.  The gauge bosons associated with the
electroweak group $SU(2)_{heavy} \times SU(2)_{light} \times U(1)_Y$
are denoted
$W^\mu_l,\ W^\mu_h$ and $X^\mu$.  The charges of the ordinary fermions
are
\eqn\fermions{\eqalign{
{\rm LH\ heavy\ (t,b)\ quarks}\ \ :&\ \ (2,1)_{1/6}\cr
{\rm LH\ light\ quarks}\ \ :&\ \ (1,2)_{1/6}\cr
{\rm LH\  leptons}\ \ :&\ \ (1,2)_{-1/2}\cr
{\rm RH\ quarks\ and\ leptons}\ \ :&\ \ (1,1)_Q\cr}}
where $Q$ is the electric charge of the right-handed
fermion~\foot{Cancellation of the $SU(2)^2 \times U(1)$ anomalies of
these fermion representations is model-dependent and will not be
addressed  here}.  The
$U(1)_{EM}$ to which the electroweak group breaks is generated by
\eqn\q{Q= T_{3l}+T_{3h}+Y.}
The photon eigenstate can be written in terms of two weak mixing
angles,
\eqn\pho{A^\mu = \st \sp \,W_{3l}^\mu + \st \cp \,W_{3h}^\mu +\ct X^\mu}
where $\theta$ is the usual weak angle and $\phi$ is an additional
one.  Eqns.~\q\ and \pho\ imply that the gauge couplings are
\eqn\eone{ \gq={e\over s\st}\,,\quad
\gl={e\over c\st}\,,\quad
\gp={e\over \ct}\,, }
where $s\equiv\sp\,$ and $c\equiv\cp$.

The order parameters of the spontaneous symmetry breaking transform as
\eqn\phii{\langle \varphi \rangle \sim (1,2)_{1/2},\ \ \ \ \ \langle
\sigma\rangle \sim (2,2)_0 \,\,\, .}
The breaking pattern shown in \lotsa\ results when they
acquire values
\eqn\phivev{ \left\langle\varphi\right\rangle=\pmatrix{
0\cr {v/{\sqrt 2}}\cr}, \ \ \ \ \ \left\langle\sigma\right\rangle=\pmatrix
{u&0\cr 0&u\cr}\,\,\,.}
In the scenario we are considering, the role of $\langle\varphi\rangle$ is
played by the TC condensate, while the physical origin of
$\langle\sigma\rangle$ is model-dependent.

While the standard electroweak model is defined in reference to three
experimental inputs, $\alpha_{EM}$, $G_F$ and $M_Z$, this extended
model requires two additional parameters (a mass scale and a coupling)
for a complete definition.  Taking these to be $u$ and $c^2$, we
can immediately place limits on them.  Equation \lotsa\
implies that the mass scale $u$ is
bounded from above by $f$, which we have already related to the size of
the top-quark mass.  Using relation \rewrite\ and defining $x \equiv
u^2/v^2$ we find
\eqn\ubound{x \approx \left({u \over f}\right)^2 {8 \pi v \over m_t}}
so that a top-quark mass of order 150 GeV implies $x$ less than or of
order 40.  The mixing angle $c^2$, on the other hand, is easily
bounded from below.  Because $g_h$ is essentially the value of the
technicolor coupling at a scale of order a TeV (the low ETC scale), we
expect it to be large compared to the weak coupling; hence we expect
 from relation \eone\ that $c^2$ will be relatively small.  However if
$c^2$ is too small, then $g_h$ will exceed the ``critical'' value at
which technifermion chiral symmetry breaking occurs. As an estimate of
this value, we use the results of the gap-equation analysis of chiral
symmetry breaking in the ``rainbow'' approximation \acref.
Then we require
\eqn\nocrit{c^2 \gae {\alpha_{EM}\over{\sin^2\theta}} {3
C_2\over{\pi}} = .03 \left({N^2 - 1 \over {2N}}\right)}
where $C_2$
is the quadratic Casimir of the technifermion representation of the
technicolor group and the last equality uses $C_2$ for the fundamental
of $SU(N)$ technicolor .

To calculate the effect of gauge-boson mixing on the $Z \to b \bar{b}$
branching ratio, we must examine the light $W$ and $Z$ states.  It is
most convenient to discuss the mass eigenstates in the rotated basis
\eqn\etwo{W^{\pm\,\mu}_1=s\,W^{\pm\,\mu}_l+c\,W^{\pm\,\mu}_h\,,\ \
W^{\pm\,\mu}_2=c\,W^{\pm\,\mu}_l-s\,W^{\pm\,\mu}_h\,}
\eqn\efiv{Z_1^\mu=\ct\,(s\,W^\mu_{3l}+c\,W^\mu_{3h})-\st\,X^\mu\,,\ \
\ Z_2^\mu=c\,W^\mu_{3l}-s\,W^\mu_{3h}\,,}
in which the gauge covariant derivatives separate neatly into standard and
non-standard pieces
\eqn\derw{\partial^\mu + ig\left( T_l^\pm + T_h^\pm \right) W^{\pm\,\mu}_1
+ig\left( {c \over s}T_l^\pm - {s \over c}T_h^\pm \right) W^{\pm\,\mu}_2,}
\eqn\derz{\partial^\mu + i{g \over {\ct}}\left( T_{3l} + T_{3h} -
\sin^2\theta \,Q \right) Z^\mu_1 +
ig\left( {c \over s}T_{3l} - {s \over c}T_{3h} \right)
Z^\mu_2.}
where $g={e \over \st}$.
The mass-squared matrix for the $Z_1$ and $Z_2$
is as follows (because the photon is massless, the $3 \times 3$ matrix for the
neutral
bosons can be reduced to a $2 \times 2$ matrix \hgsw .)
\eqn\esev{M_Z^2=\left({e v \over {2 \st}} \right)^2\,
\pmatrix{{1\over \cos^2\theta}&
{- s\over c\cos\theta}\cr{- s\over c\cos\theta}&
{x\over s^2c^2}+{s^2\over c^2}\cr}.}
The mass-squared matrix for the $W_1$ and $W_2$ may be obtained by
simply setting $\ct = 1$ in the above matrix.

We expect $x$ to be reasonably large and we
diagonalize the $W$ and $Z$ mass matrices in the limit of large $x$;
we will find that this is self-consistent.  The perturbative
expressions we obtain for the light $W$ and $Z$ masses and eigenstates
to leading order in ${1\over x}$ are\foot{An analysis of the
low-energy charged currents, analogous to that carried out in
ref. \ununify , shows that $\sqrt{2} G_F = {1\over v^2}
+ {1\over u^2}$.}
\eqn\lw{(M_W^L)^2 \approx
\left({\pi \alpha_{EM} \over{\sqrt{2} G_F \sin^2\theta}}\right)
\left(1 + {1\over x}(1 - s^4)\right),\ \ \ \
W^L\approx W_1+{c s^3 \over {x}}\,W_2}
\eqn\lz{(M_Z^L)^2 \approx
\left({\pi \alpha_{EM} \over{\sqrt{2} G_F \sin^2\theta\cos^2\theta}}\right)
\left(1 + {1\over x}(1 - s^4)\right),\ \ \ \
Z^L\approx Z_1+{c s^3 \over {x\cos\theta}}\,Z_2}
Note that as $c^2 \to 0$ the light $W$ and $Z$ approximate the $W_1$ and
$Z_1$ states which have standard couplings to fermions.

Now we can compute the effect of $Z_1 - Z_2$ mixing on $R_b$. From
\derz\ and \lz\ we find a change in the coupling of the $Z$ to
$b$-quarks
\eqn\dgmix{
\delta g_L \approx {e \over \sin\theta \cos\theta}
{s^4 \over 2x}\ \ .}
This results in the following change in
$\Gamma_b$ due to the gauge boson mixing
\eqn\some{ {\delta \Gamma_b \over {\Gamma_b}} = {2 g^b_L \delta g^b_L
\over {(g^b_L)^2 + (g^b_R)^2}} \approx -2.3 {s^4\over{x}}\ \ .}
The effect on all quarks other than the $b$ may be computed
similarly. Equations \lz\ and \derz\ imply a shift in
the left-handed couplings of these quarks in the amount
\eqn\otherb{
\delta g_L = {e\over \sin\theta \cos\theta} {c^2 s^2\over x} T_3\ \ ,}
where $T_3$ is $+{1\over 2}$ for up-quarks and $-{1\over 2}$ for
down-quarks.  This results in a change in the $Z$ width
to hadrons other than the $b$ by
\eqn\somtwo{
{\delta \Gamma_{h\neq b} \over \Gamma_{h\neq b}}
\approx + 2.3 {s^2 c^2 \over x}\ \ .}
Combining \some\ and \somtwo\ and using \ubound, we find that the effect of
gauge-boson mixing on $R_b$ is
\eqn\finally{
{\delta R_b \over R_b} \approx (1-R_b) \left(
{\delta \Gamma_b \over \Gamma_b} -
{\delta \Gamma_{h\neq b} \over \Gamma_{h\neq b}}\right) \approx
-4.4\%\ s^2
\left({f\over u}\right)^2 \left({m_t \over 150{\rm GeV}}\right)
\ \ .}

We see that the effect of gauge-boson mixing on the $Z \to b \bar{b}$ decay
width is of the same order of magnitude, but opposite sign, to the
effect due to the ETC dynamics of top-quark mass generation, eqn.
\shiftr . Unlike the case of commuting ETC models, we cannot make a
model-independent statement of the overall size or sign of the change
in the $Z\to b\bar{b}$ branching ratio: the overall effect may be small
and may even {\it increase} the $Z b \bar{b}$ branching ratio.

Finally, we should mention that gauge-boson mixing leads to a number
of other potentially large effects as well. For example, because the
$Z$ mass is used as an input in fits of LEP results to the standard
model, the shift in the tree-level $Z$ mass in eqn. \lz\ results in a
redefinition of $\sin^2 \theta$. Both this effect and the direct shift
in the $W$ mass (eqn. \lw ) affect $M_W$. Furthermore, there are also
changes in the coupling of the $Z$ to leptons of the same form as
those in eqn. \otherb . None of these effects change our computation
of $R_b$, and all vanish in the limit that $c^2 \to 0$. However, when
combined with the model-dependent radiative corrections due to the
technicolor sector \tcew , they will provide further constraints
\terning\ on $c^2$ and $f/u$.

\newsec{A Non-Commuting Extended Technicolor Scenario} So far we have
discussed the possibility of the TC and weak gauge groups' being
embedded in the ETC gauge group, but we have avoided any mention of
how the fermion representations of these subgroups are embedded in
representations of ETC.  This is a complicated problem, and we will
satisfy ourselves here with a sketch of how this might be done: we
will only discuss the generation of the mass of the top quark and will
ignore questions about anomalies.  What we have in mind is a scenario
where an $SU(N+2)_{ETC}\otimes SU(2)_{light}$ gauge group breaks to
$SU(N)_{TC} \otimes SU(2)_{heavy} \otimes SU(2)_{light}$.  We imagine
that the ETC model contains the following representations (among
others) of $SU(N+2)_{ETC} \otimes SU(3)_C \otimes SU(2)_{light}$:
\eqn\etcreps{\eqalign{ ({\bf N + 2},{\bf 3}, {\bf 1}), & \cr ({\bf
\overline{A_{N+2}}},{\bf \overline{3}},{\bf 1}) & ,\cr }} where ${\bf
A_{N+2}}$ is the antisymmetric tensor representation of $SU(N+2)$ with
dimension \eqn\dim{ d({\bf A_{N+2}}) = {(N+1)(N+2)\over 2}~.  } When
$SU(N+2)_{ETC}$ breaks, we have the following representations under
$SU(N)_{TC} \otimes SU(2)_{heavy} \otimes SU(3)_C \otimes
SU(2)_{light}$: \eqn\tcreps{\eqalign{ ({\bf 1},{\bf 2},{\bf 3}, {\bf
1})\ , \ ({\bf N },{\bf 1},{\bf 3}, {\bf 1}) & \cr (t,b)_L\ \ \ \ \ \
, \ \ U_L\ \ \ \ \ \ \ \ \ & \cr ({\bf 1},{\bf 1},{\bf
\overline{3}},{\bf 1})\ ,\ ({\bf \overline{A_{N}}},{\bf 1},{\bf
\overline{3}},{\bf 1})\ ,\ ({\bf \overline{N}},{\bf 2},{\bf
\overline{3}},{\bf 1}) & \cr t_R^c\ \ \ \ \ \ \ \ \ \ , \ \ A_R^c\ \ \
\ \ \ \ \ \ \ \ \ , \ (U,D)_R^c \ \ \ \ & ~. \cr }} After
$SU(2)_{heavy}$ and $SU(2)_{light}$ mix, we have the following
representations under $SU(N)_{TC} \otimes SU(3)_C \otimes SU(2)_{L}$:
\eqn\finalreps{\eqalign{ ({\bf 1},{\bf 3}, {\bf 2})\ ,\ ({\bf N },{\bf
3}, {\bf 1})&\cr (t,b)_L \, \ \ \ \, ,\,U_L \ \ \ \ \ \ \ &\cr ({\bf
1},{\bf \overline{3}},{\bf 1})\ ,\ ({\bf \overline{A_{N}}},{\bf
\overline{3}},{\bf 1})\ ,\ ({\bf \overline{N}},{\bf \overline{3}},{\bf
2})&\cr t_R^c\,\,\,\,\,\ \ \ \ \ ,\ \ A_R^c \ \ \ \ \ \ \ \,\,\,\,\,,\
(U,D)_R^c&~.  }}

When $SU(N)_{TC}$ gets strong the $(U,D)_R^c$ doublet
condenses with the $U_L$ (and with a $D_L$ which we have not
described), breaking $SU(2)_L$.
The ETC gauge bosons with masses of order $g_h f /2$ in this scenario
have the following quantum numbers under $SU(N)_{TC} \otimes SU(3)_C
\otimes SU(2)_{L}$:
\eqn\etcgb{\eqalign{
({\bf  N},{\bf 1}, {\bf 2})\ ,\ ({\bf \overline{ N} },{\bf 1}, {\bf 2})\ ,\
({\bf  1},{\bf 1}, {\bf 1})&~.\cr
}}
Exchange of the weak-doublet ETC gauge bosons gives rise
to the operator \tmasop\ with $\xi=1$.


\newsec{Conclusions}

We have explored the possible effects of a new type of ETC model on
the $Z$ partial decay width into $b \overline{b}$.  Two types of
effects were uncovered: a direct vertex correction due to exchange of
the ETC gauge boson responsible for top-quark mass generation, and a
correction due to mixing of the $Z$ with a techni-neutral ETC gauge
boson ($Z_2$).  The former effect increases the $Z$ partial width,
while the latter decreases it.

Unlike the case of commuting ETC models, we cannot make a
model-independent statement of the overall size or sign of the change
in the $Zb\bar{b}$ branching ratio: the overall effect may be small
and may even {\it increase} the $Z b \bar{b}$ branching ratio.  It
will be important to explore this class of models further, since
experiments at LEP find that the $Zb\bar{b}$ width lies slightly above
the standard model prediction.

\centerline{\bf Acknowledgments}

We thank Ken Lane for discussions and for reading the manuscript.
We appreciate the hospitality of the Institute for Theoretical Physics, where
part of this work was completed.
R.S.C. acknowledges the support of an Alfred P. Sloan Foundation
Fellowship, an NSF Presidential Young Investigator Award, an SSC
Faculty Fellowship from the Texas National Research Laboratory
Commission, and a DOE Outstanding Junior Investigator Award.  EHS
acknowledges the support of an American Fellowship from the American
Association of University Women.
{\it This work was supported in part by the National Science
Foundation under grant PHY-9057173, by the Department of Energy under
contract DE-FG02-91ER40676, and by the Texas National Research
Laboratory Commission under grant RGFY92B6.}

\listrefs
\bye